\documentclass[onecolumn]{mn2e}
\usepackage{epsfig}

\title[Universal Galaxy Rotation and Dark Matter]
{\bf THE UNIVERSAL ROTATION CURVE OF SPIRAL GALAXIES.
II THE DARK MATTER DISTRIBUTION OUT TO THE VIRIAL RADIUS.}
\author[ P.Salucci ]
{P. Salucci$^{1}$, A. Lapi$^{1}$, C. Tonini$^{1}$, G. Gentile$^{2}$,
I. Yegorova$^{1}$, U. Klein$^{3}$ \\
$^1$SISSA  International School for Advanced Studies, via Beirut 4,
I--34013 Trieste, Italy \\ $^2$ University of New Mexico, Department
of Physics and Astronomy, 800 Yale Blvd NE, Albuquerque, NM 87131, USA \\
$^3$ Argelander-Institut f\"ur Astronomie,
Auf dem H\"ugel 71, D--53121 Bonn, Germany }

\begin{document}

\maketitle

\def\mincir{\raise -2.truept\hbox{\rlap{\hbox{$\sim$}}\raise5.truept
\hbox{$<$}\ }}
\def  \magcir{\raise -2.truept\hbox{\rlap{\hbox{$\sim$}}\raise5.truept
\hbox{$>$}\ }}
\def\bibitem{}{\par\noindent\hangindent 20pt}

\begin{abstract}
In the current $\Lambda$CDM cosmological scenario, $N$-body
simulations provide us with a Universal mass profile, and
consequently a Universal equilibrium circular velocity of the
virialized objects, as galaxies. In this paper we obtain,  by
combining kinematical data of their inner regions with global
observational properties, the Universal Rotation Curve (URC) of disk
galaxies and the corresponding mass distribution out to their virial
radius. This curve extends the results of Paper~I, concerning the
inner luminous regions of Sb-Im spirals, out to the edge of the
galaxy halos.
\end{abstract}

\section{Introduction}

Rotation curves (hereafter RCs) of disk galaxies do not show any
Keplerian fall-off and do not match the distribution of the stellar
(plus gaseous) matter. As a most natural explanation, this implies
an additional invisible mass component (Rubin et al. 1980; Bosma
1981, Persic \& Salucci, 1988) that becomes progressively more
conspicuous for the less luminous galaxies (e.g.: Persic \& Salucci
1988, 1990; Broeils 1992a). Moreover, the kinematical properties of
Sb-Im spirals lead to the concept of the {\it Universal Rotation
Curve} (URC)  implicit in  Rubin 1985, pioneered in Persic and
Salucci, 1991 and set in  Persic, Salucci \& Stel (1996, hereafter
PSS, Paper I): RCs can be generally represented out to $R_{l} $, the
outermost radius where data are available,  by $V_{URC}(R; P)$, i.e.
by a {\it universal} function of radius, tuned by some galaxy
property $P$. $P$ can be a global property such as the luminosity
and the disk or halo mass or a well defined local quantity like
$V_{opt}$. In any case it serves as the galaxy identifier. In PSS
individual RCs and a number of coadded RCs proved the URC paradigm
being well fitted by an analytical  Curve, $V_{URC}(r/R_{opt},
L)$\footnote{The reader is directed to  PSS for the details of the
procedure.}, a function which is the sum in quadrature of  two
terms:  $V_{URCD}$ and $V_{URCH}$, each representing the disk or
halo contribution to the circular velocity:
$$
V^2_{URC} = V^2_{URCD} + V^2_{URCH}
\eqno(1)
$$
The stellar component was described by  a Freeman disk (Freeman
1970) of surface density $\Sigma_D(r) = {M_D \over 2 \pi \ R_D^2}\,
e^{-r/R_D}$ and  contributing  to the circular velocity $V$ as:
$$
V^2_{URCD}(x) = {1\over 2} {GM_D\over {R_D}} (3.2 x)^2 (I_0K_0-I_1K_1)
\eqno(2a)
$$
where $x=r/R_{opt}$ \footnote{We define the "disk size" $R_{opt} \equiv
3.2\,R_D$} and $I_n$ and $K_n$ are the modified Bessel
functions computed at $1.6~x $. The  dark matter component (with
$V_{URCH}^2 (r) = {G\, M_{H}(<r)\over r}~$) was described by means
of a  simple halo velocity profile :
$$
V_{URCH}^2(x) = {1\over {4 \pi}} G \rho_0 a^2  x^2/(a^2 +x^2)
\eqno(2b)
$$
The above implies a density profile with an inner flat velocity
core of size $\sim a R_{opt}$, a central density $\rho_0$, an outer
$r^{-2}$ decline. The sum of the contributions (2a) and (2b) well
fit all the PSS data with $\rho_0$, $a^2$ specific functions of
luminosity (see PSS). Let us remind that  disk masses $M_D$ of
spirals were found in the range $10^9~$M$_\odot \leq M_D \leq 2
\times 10^{11}$~M$_\odot$.

The URC for the purpose of this work matches well the individual
RCs  of late type spirals (see also Appendix for a discussion).  
It is useful to express the  URC paradigm in the following way: at any chosen
radius, the URC predicts the circular velocity of a (late type)
spiral of {\it  known} luminosity and disk scale-length,  within an
error that is one order of magnitude smaller than the variations
it shows i)  at different radii and ii) at any radius, with respect
to objects of different luminosity.

Let us remind that the Universal curve built in PSS holds out to
$R_l$, uses the luminosity as the galaxy identifier and the disk
scale-length as a unit of measure for the radial coordinate. We will
label it as URC${_0}$ to indicate it as the first  step of a
definitive  function  of the dark radial coordinate, able to
reproduce  the observed RCs  of   spirals. URC$_0$ provides
fundamental knowledge on the mass distribution in spirals, while it
suffers from three  limitations: 1) it strictly  holds only in a
region extended less than $5\%$ the DM halo size (see below) 2) the
velocity  profile of the halo component, valid out to $R_l$, cannot
be extrapolated to radii of cosmological interest 3) it identifies
objects by  their luminosities, rather than by  their virial masses.
Let us point out that the URC$_0$ has been often and  successfully
used as an observational benchmark for theories, but this, only for
$R<R_l$ and after that  a relation between the halo mass and the
galaxy  luminosity was assumed.

On the other side, high--resolution cosmological N--body simulations
have shown that, within the ($\Lambda$) Cold Dark Matter (CDM) scenario,
dark halos achieve a specific equilibrium density profile characterized
by a universal shape and, in turn, an universal halo circular
velocity (Navarro, Frenk \& White, 1997,
NFW), $V_{\rm NFW}(R, M_{vir})$ in which  the virial mass $M_{vir}$ and virial
radius $R_{vir}$ are  the galaxy identifier and radial coordinate.
$$
\rho_H (r)= {M_{vir}\over 4\pi R_{vir}^3}\, {c^2\,g(c)\over x\,(1+cx)^2}~,
\eqno(3a)
$$
where $x\equiv r/R_{vir}$ is the radial coordinate, $c$ is the concentration
parameter, and $g(c)= [\ln(1+c)-c/(1+c)]^{-1}$. The parameter
$c$ is found to be a weak function of the halo mass, given by
$c\approx 14\, \left({M_{vir} / 10^{11}\, \rm M_{\odot}}
\right)^{-0.13}$ (Bullock et al, 2001, Dutton, 2006, Gnedin 2006).
This leads to
$$
V_{\rm NFW}^2(r)= V_{vir}^2 \frac{c}{g(c)} \frac {g(x)}{x}~,
\eqno(3b)
$$
with $V_{vir}=V(R_{vir})$. It is interesting to note that in this
scenario the present--day circular velocity, which also includes a
baryonic component arranged in a disk, is predicted to be a {\it
Universal} function of radius, tuned by few galaxy parameters (Mo,
Mao \& White 1998). However, it is well known that observations of
spiral galaxies favor density concentrations lower than those
predicted for CDM by Eq.~(3a): DM halos detected around spirals do
not show the  NFW central cusp in favor of a core-like structure
(Gentile et 2005; van den Bosch \& Swaters 2001; Swaters et al.
2003; Weldrake et al. 2003; Simon et al. 2005; Donato et al, 2004;
Gentile et al. 2007).

Therefore, the reconstruction of the mass distribution of DM halos
from observations {\it in parallel}  with that emerging from N-body
simulations is required not only as a normal scientific routine, but
also in view of a theory-vs-observations likely   disagreement.

As an alternative to the simulation method, we will support the URC
paradigm by means of a set of proper observational data and we will
derive an analytical form for this curve, valid from the galaxy
center out to its virial radius and characterized by the halo mass
as the galaxy identifier. In detail, we extend/improve the URC$_0$
in PSS a) by adopting a different halo profile, proper to describe
the halo distribution out to the virial radius, b) by using a number
of RCs substantially more extended than those in PSS
and c) by exploiting the relationship between the disk
mass $M_D$, and the virial galaxy mass $M_{vir}$, recently obtained
by Shankar et al. (2006). This will allow to build an
"observational" Universal Curve, $V_{URC}(R; M_{vir})$, extended out
to $R_{vir}$ and having the virial mass as the galaxy identifier.
This curve is  the observational counterpart of the universal
$\Lambda$CDM  NFW N-body generated profile.

\begin{figure}
\centerline{\psfig{figure=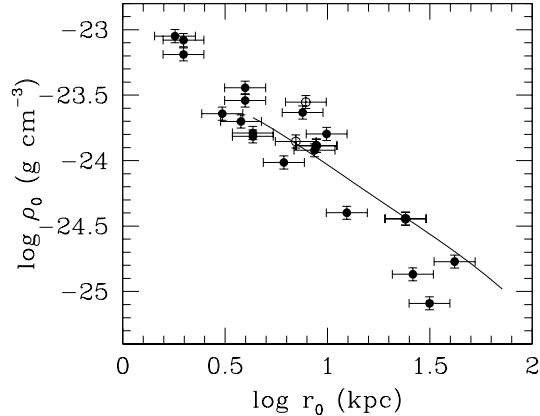,height=7.5cm}} \vskip -0.3truecm
\caption{The core radius vs. virial mass relations for
the SE sample  and  the present work (solid line).}
\vskip -0.2truecm
\end{figure}
 
While  pointing that the concept behind the   Universal
Rotation Curve may be  valid also for  galaxies of  different
Hubble Types (see Salucci and Persic, 1997), but
a number of issues  are still open and will be dealt elsewhere:

i) Sa galaxies amount, by number, to less than 10\% of  the whole
spiral population, and are important  objects  in view
of the dual nature of their stellar distribution. They show RC
profiles  with  a  clear systematics with luminosity (Rubin et al. ,
1985), but, not unexpectedly, with some difference from those of
the URC$_{0}$ (Noordmeer, 2007).

ii) Dwarf spirals with $V_{opt}<50 km/s $ are not well studied and
included in the URC yet, also because in these objects  the  RCs  do
not coincide with  the circular velocity, being  significant the
complex asymmetric drift correction.

iii) The kinematical properties of spirals of very high  stellar disk
mass are not presently investigated with a suitably large sample.

iv) A possible additional URC physical parameter 
(e.g.  the surface stellar density) to take
care of the (small) variance of the RCs profiles that seems to be
unaccounted by the luminosity.

Finally, let us remind that, in a flat cosmology with matter
density parameter $\Omega_M = 0.27$ and Hubble constant $H_0 = 71~
\mathrm{km~s}^{-1}~\mathrm{Mpc}^{-1}$, at the present time,
the halo virial radius
$R_{vir}$, i.e. the size of the virialized cosmological perturbation
of mass $M_{vir}$ scales with the latter as:
$$
R_{vir} = 259 \Big( {M_{vir}\over {10^{12} \ \rm M_\odot}
}\Big)^{1/3} \ \rm kpc~. \eqno(4)
$$
(see e.g. Eke et al. 1996).

\section{The Universal halo velocity profile}

We assume that the DM mass distribution is described by the Burkert
(1995) profile
$$
\rho (r)={\rho_0\, r_0^3 \over (r+r_0)\,(r^2+r_0^2)}~, \eqno(5a)
$$
where $r_0$ is the core radius and $\rho_0$ the effective core
density, in principle two independent parameters. Correspondingly,
the total halo mass inside radius $r$ is given by $M_{H}(r) = 4 \,
M_0 \, \left[\ln\left( 1+\frac{r}{r_0} \right) -
\tan^{-1}\left(\frac{r}{r_0} \right) +
\frac{1}{2}\,\ln\left(1+\frac{r^2}{r_0^2}\right) \right] $ with
$M_0=1.6\,\rho_0\,r_0^3$, so that:
$$
V^2_{URCH}(r)= 6.4 \ G \ {\rho_0 r_0^3\over r} \Big\{ ln \Big( 1 +
\frac{r}{r_0} \Big) - \tan^{-1} \Big( \frac{r}{r_0} \Big) +{1\over
{2}} ln \Big[ 1 +\Big(\frac{r}{r_0} \Big)^2 \Big] \Big\}~. \eqno(5b)
$$
Inside $R_l$ this profile is indistinguishable from the halo term
(2b) in the URC$_0$ (Salucci \& Burkert, 2000, Gentile et al 2004).
At larger radii, the mass diverges only logarithmically with radius
and converges to the NFW velocity profile, provided that $r_0 \ll
R_{vir}$.

We fit the set of individual and coadded RCs of PSS with $V_{URC}(R; M_D,
\rho_0, r_0)$ and derive the model parameters $M_D$, $\rho_0 $, $r_0$
(see  Salucci and Burkert,  2001):
$$
\log {\rho_0\over \mathrm{g}~\mathrm{cm}^{-3}} = -23.515-0.964\,
\left({M_D}\over 10^{11}\, \rm M_{\odot}\right)^{0.31}~
\eqno(6a)
$$
and
$$
\rho_0 = 5\times 10^{-24}r_0^{-2/3}e^{-(r_0/27)^2} \rm g~cm^{-3}~.
\eqno(6b)
$$
Eqs.~(1)-(2a)-(5b)-(6) define the URC out to $R_l$, $V_{URC}(R, M_D,
r_0)$, from the "baryonic perspective". Let us notice that, as
result of the RC mass modeling, $r_0$, differently from $M_D$ and
$\rho_0$, has quite large fitting uncertainties, viz. $\delta
r_0/r_0 = 0.3 - 0.5$. Following our empirical approach, we do not
extrapolate the URCH determined inside $R_l$ out to $R_{vir} \gg
R_l$. In that this will be uncertain besides of unknown validity.
This quantity will be derived in the next section.

\section{The URC out to the virial radius}

We overcome the two main limitations of the  URC$_{0}$, its
problematic extrapolation between $R_l$ and $R_{vir}$ and the
uncertainty in the estimate of the core radius, by determining the
latter by means of a new {\it outer} observational quantity, the
halo virial velocity $V_{vir}\equiv [G
M_{vir}/R_{vir}(M_{vir})]^{1/2}$, related to the virial mass through
Eq.~(4). In detail, we obtain $V_{vir}$ from the disk mass, suitably
measured from inner kinematics through its relationship to the
virial mass found by Shankar et al. (2006) that is  well represented (i.e. within $5 \%$,  
the actual relation  we use is shown in Fig. 2 and given in the Code indicated in the Discussion) by:
$$
M_D = 2.3 \times 10^{10} \ M_\odot \frac {[M_{vir}/(3 \ 10^{11} \
\rm M_\odot) ]^{3.1} }{1+[M_{vir}/(3 \ 10^{11} \rm M_\odot)]^{2.2}} ~.
\eqno(7)
$$
This relationship is a consequence of the existence of i) the
universal stellar mass function, (Bauldry et al 2004, Bell et al  2003)
 and of ii) the
cosmological halo mass function as indicated by $N$-body
$\Lambda$CDM simulations .\footnote {There is no inconsistency in
adopting the $\Lambda$CDM halo mass function and cored halo mass
models, in that the latter can be formed astrophysical from the
cosmological cuspy ones. Since both functions account the same
cosmological objects, the Jacobian of their transformation defines a
relation between the disk and virial mass in spirals (see Shankar et
al. 2006).}  Let us notice that this relationship is obtained
without assuming any halo density profile, so that it can be
combined with the mass modelling of the inner kinematics.

Let us first derive $R_D(M_{vir})$,  the disk scale-length  as a
function of the halo mass,   by inserting Eq.~(7) in the
relationship:
$$
\log {{R_D}\over \mathrm{kpc}} = 0.633+0.379\,\log{M_D\over
10^{11}\, \rm M_{\odot}}+0.069\, \left(\log{M_D\over 10^{11}\,
\rm M_{\odot}}\right)^2
\eqno(8)
$$
obtained  in  PSS. We note that no result
of this work is affected by the observational uncertainties on the
relationship Eq.~(8).

It is worth to compute the radial extrapolation needed to reach $R_{vir}$
from $R_l= 6 R_D$, a quantity that can also be associated with the baryonic
collapse factor $F= R_{vir}/R_D$; we find $F \approx 90 - 15 \ log {M_{vir}
\over {10^{11} M_\odot}}$, i.e. about 25 times the disk size $\sim 3 R_D$.

Eq.~(7) in combination with Eq.~(4) allows us to add,
 to the PSS set of kinematical data leading to the
URC$_0$, a new  observational quantity: $V_{vir}(M_D)
=GM_{vir}/R_{vir}$, relative to the virial radius.  Then, we
determine  the core radius not from the inner kinematics, but as
the value of $r_0$  for which the velocity model described by
Eqs.~(1), (2a), (5b) and (6a), matches (at $R_{vir}$) the virial
velocity  $V_{vir}$ given by Eq.~(7) and (4). Let us write this as:
$$
{GM_{vir} \over {R_{vir}(M_{vir})}}=   V^2_{URCH}
[R_{vir}(M_{vir}); \rho_0(M_{vir}), r_0)] \eqno(9a)
$$
where $\rho(M_{vir})$ as a short form for $\rho(M_D(M_{vir}))$ with
Eq.~(7) inserted in Eq.~(6a). From the PSS inner kinematics we get
the values of $\rho_0$ and $M_D$ according to Eq.~(6a), so that
Eq.~(9a) becomes an implicit relation between $r_0$ and $M_{vir}$
($c_1, c_2$ are known numerical constants):
$$
M_{vir} = c_1 {\rho_0(M_{vir}) r_0^3\over {c_2 M_{vir}^{1/3}}}
\Big\{ ln \Big( 1 + \frac{c_2 M_{vir}^{1/3} }{r_0} \Big) -\tan^{-1}
\Big( \frac{c_2 M_{vir}^{1/3}}{r_0} \Big) +{1\over {2}} ln \Big[ 1
+\Big(\frac{c_2 M_{vir}^{1/3}}{r_0} \Big)^2 \Big] \Big\} ~.
\eqno(9b)
$$
The above can be numerically solved for any $M_{vir}$, and the
solution can be approximated by:
$$
\log{(r_0/\mathrm{kpc})} \simeq 0.66+0.58\,
\log{(M_{vir}/10^{11}\, \rm M_{\odot})} ~.
\eqno(10)
$$
(an higher order  approximation is given in the Code indicated in the Discussion).
Let us stress that the present derivation of $r_0$  is very solid
with respect to observational uncertainties: errors up to a factor 2
in $M_{vir}$ in eq (7) trigger errors in $r_0$ lower than $40\%$,
and errors in the outer halo velocity slope ($ 0.1 \leq R/ R_{vir}
\leq 1 $) lower than $0.1$. This is  certainly smaller than the
scatter of  values with which this quantity is found by $N$-Body
simulations and by SPH/semi analytical studies of galaxy formation
including the baryonic components.

\begin{figure}
\centerline{\psfig{figure=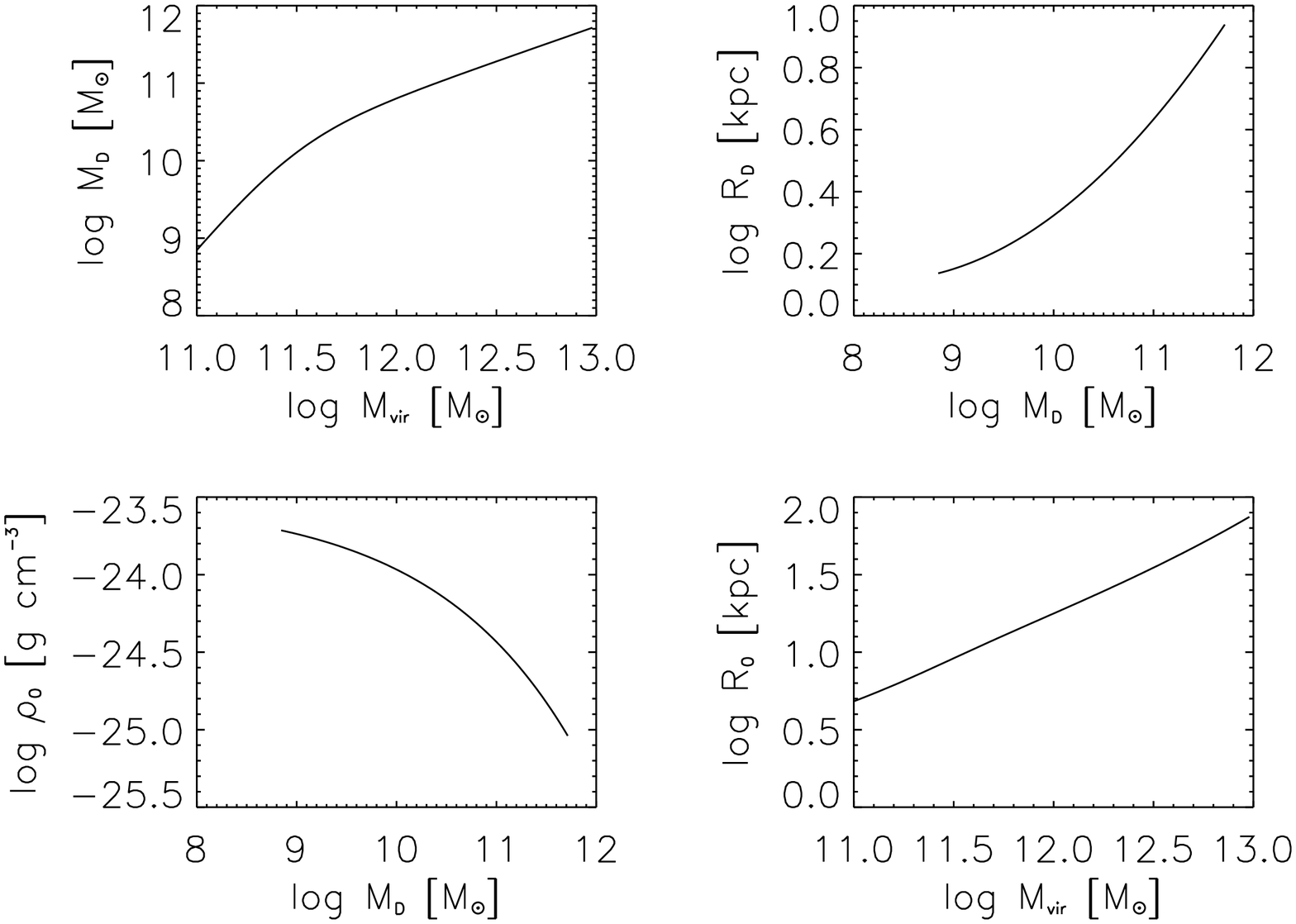,height=7.5cm}} \vskip -0.3truecm
\caption{The various relationships used in this paper.} \vskip
-0.2truecm
\end{figure}

It is worth investigating a number of recently published
super-extended (SE)  RCs  (Donato et al. 2005, Gentile et al 2004,
Salucci et al 2003). They  reach  a radius  larger  than  $5\%$ (and
up to  $15\%$) of the virial radius, i.e. a radius  at least  twice
as extended  as  those of  the  sinthetic curves in  PSS. The mass
modelling of these  SE  RCs    (made in the original papers) shows
an $r_0$ vs. $M_{vir}$ relationship that is in good agreement with
Eq.~(10). Relation (10) and the above individual  values differ by
$20\%$ -$40 \%$ from those determined  from the inner kinematics
alone and given by means of  eq (6b). Since in this paper (also
because $R_l\sim r_0$), eq (6) is considered a prediction of the
inner mass modelling rather than an actual  measurement,  such  good
agreement indicates the soundness of the PSS  mass modelling.

Let us notice that  only for a range of values of the  crucial
 quantity $V(R_l)$-  $V_{vir}$, with the first term obtained by
the inner kinematics and the second one via Eqs.~(4) and (7), there
is a solution for eq.(9b), therefore, the existence of eq (10) and
the  agreement of the values of the Burkert core radii, measured
independently at $0.05 R_{vir} $, $0.1 R_{vir} $ and $R_{vir}$ are
important tests passed by this profile.

Then, by means of Eqs.~(1), (2a)-(5b)-(6a)-(10), we construct the
full URC, extended out to the virial radius and with the virial mass
as the galaxy indicator. It is useful to show the relationships we
use (see Fig.~2). The mass model includes a Burkert DM halo of
central density $\rho_0$, of core radius of size $r_0$ and a Freeman
disk of mass $M_D$. The URC fits nicely the available velocity data
out to $R_l$ and it is valid out to the virial radius, where it
exactly matches $V_{vir}$. Moreover, since $M_{vir}$ is the quantity
that in theoretical studies identifies a galaxy, we overcome the
main limitation of URC$_0$.

We consider all of the three coordinate systems $r, r/R_D,
r/R_{vir}$ equivalent to represent the main structural properties of
the mass distribution in spirals, but each of them showing  some
particular aspects. More specifically, it is then possible and
useful  to build several "URCs" i.e. $V_{URC}(r/R_{coo}; P)$, where
$P$ is a galaxy identifier ($M_D, M_{vir}, L$) and $R_{coo}$ a
radial coordinate ($r, r/R_D, r/R_{vir}$). Although not all these
URCs are independent in a statistical sense, they all are relevant
in that they all well reproduce the individual RCs and each of them
highlights particular properties of the mass distribution.

\begin{figure}
\centerline{\psfig{figure=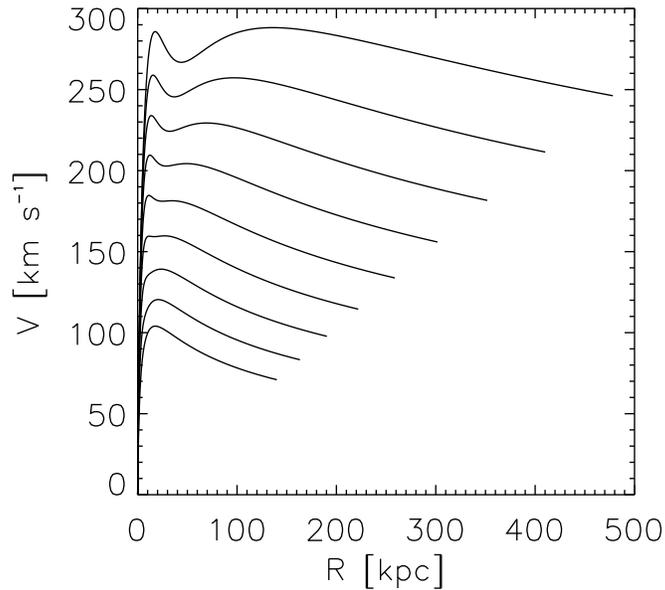,height=9cm}} \vskip -0.2truecm
\caption{The Universal Rotation Curve in physical units. Each curve
corresponds to $M_{vir}=10^{11} \ 10^{n/5} \rm M_\odot $, with $n =
1 \ldots 9$ from the lowest to the highest curve.} \vskip -0.5truecm
\end{figure}

In Fig.~3 we show $V_{URC}(r; M_{vir})$, the URC in physical units
with the objects identified by the halo virial mass; each line
refers to a given halo mass in the range $10^{11}\, \rm M_{\odot}\la
M_{vir} \la 10^{13}\, \rm M_{\odot}$;  the halo mass determines both
the amplitude and the shape of the curve. Note the contribution of
the baryonic component, negligible for small masses but increasingly
important in the larger structures, mirrors the behavior of the
$M_{vir}-M_D$ relation. The general existence of an inner peak is
evident but, especially at low masses, it is due to both dark and
stellar components. Remarkably, the maximum value of the circular
velocity occurs at about $15 \pm 3$~kpc, independent of the galaxy
mass: this seems to be a main kinematic imprint of the DM - luminous
mass interaction occurring in spirals. Furthermore, Fig.~3 shows
that the "Cosmic Conspiracy" paradigm has no observational support:
there is no fine tuning between the dark and the stellar structural
parameters to produce the same particular RC profile in all objects
(e.g. a flat one). Conversely, a number of relationships between the
various structural parameters  produce a variety of RC profiles.
Moreover, the peak velocity of the stellar component
$V_{disk}^{peak} = V_D(2.2 R_D) = G M_D/R_D\ k$, with $k =const$, is
not a constant fraction of the virial velocity as is found in
ellipticals, (i.e $\sigma \propto V_{vir}$), but it ranges between
the values $1$ and $2$ depending on the halo mass.

Moreover, as in the NFW (and Burkert) RC profiles, the URC profiles
are found (moderately) decreasing over most of the halo radial
extent. The paradigm of flat rotation curves is obviously incorrect
even/especially intended as an asymptotic behavior at large radii.
In fact, we find that both $V(0.05 R_{vir})$,the velocity at the
farthest radius with available kinematical in PSS and $V(3R_D)$, a
main  reference velocity of the luminous regions of spirals, are
significantly ($10\%-30\%$) higher than the (observational) value of
$V_{vir}$. This rules out a $V = constant$ extrapolation of the
inner RCs out to regions non mapped by the kinematics and DM
dominated regions. We note that this result is independent of the
adopted halo density profile and is far from being granted on
theoretical grounds.

\begin{figure}
\vskip -.3truecm
\centerline{\psfig{figure=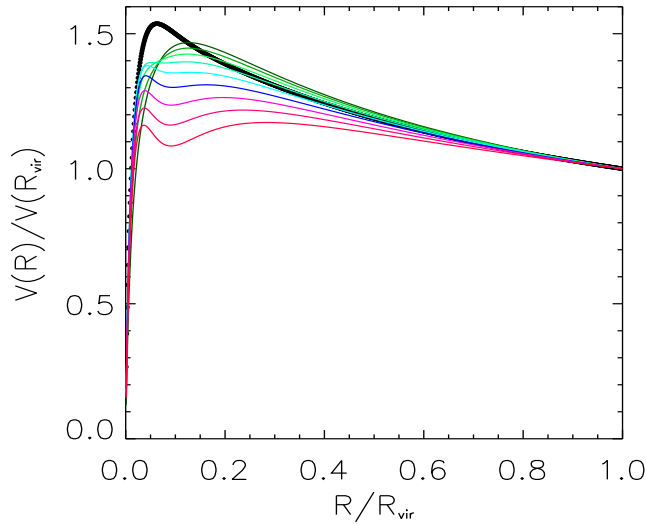,height=8cm,width=9cm }} \vskip
-0.2truecm \caption{The Universal Rotation Curve, normalized at its
virial value $V_{URC}(R_{vir})$, as a function of normalized dark
radius $ x \equiv R/R_{vir}$. Each curve,  from the highest  to the
lowest, corresponds to  $M_{vir}$ defined as in Fig.~3. The bold
line is the NFW velocity profile (see text). } \vskip -0.3truecm
\end{figure}

In Fig.~4 we frame the URC from a full DM perspective by plotting
$V_{URC}(R/R_{vir}; M_{vir})$. We set the virial mass $M_{vir}$ as
the galaxy identifier and $R/R_{vir}$ as the radial "dark"
coordinate, thus normalizing the amplitudes by $V_{vir} \propto
M_{vir}^{1/3}$. This ensemble of curves, a main goal of the present
work, is parallel to those emerging in $N$-body simulations and aims
to represent the actual velocity profiles of spirals. In these
variables the DM halos are self-similar; the whole system is
self-similar in the outer regions, while in the innermost $30\%$ of
the halo size the baryons have influenced the dynamics and broken
the self-similarity. In these coordinates it easily emerges that the
maximum of the RC occurs at very different radii, viz. at $\simeq 2
R_D$ for the most massive objects and at $\sim 10 R_D$ for the least
massive ones. Then, no reference circular velocities, to be
considered as the actual physical counterparts of the empirical
velocities of the Tully-Fisher relationship, exist in actual galaxy
rotation curves.

\begin{figure}
\centerline{\psfig{figure=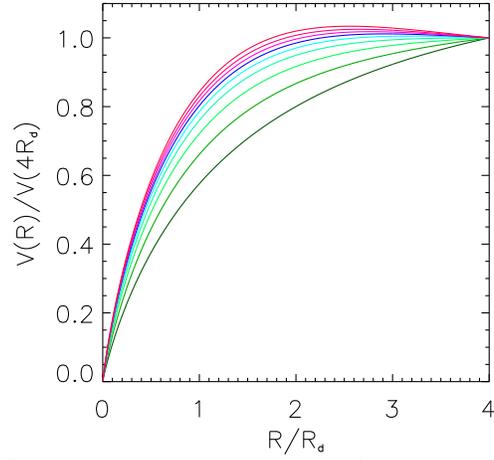,height=7cm}} \vskip -0.5
truecm \caption{The inner Universal Rotation Curve, normalized at
its value at $4R_D$, as a function of normalized stellar radius
$R/R_D$ for galaxies with  $M_{vir}$ as in Fig.~3.} \vskip 0.2truecm
\end{figure}

In Fig.~5 we zoom into the URC to look for the inner (luminous)
regions of spirals from a baryonic perspective: the URC is so
expressed as a function of the "baryonic" radial coordinate $r/R_D$.
This figure corresponds to Fig.~4 of PSS, with the important
difference that here the virial mass,  rather than the galaxy
luminosity, is the galaxy identifier. Plainly, an inverse
correlation between the average steepness of the RC slope and the
halo mass holds, similar to the slope-luminosity relationship found
by Persic \& Salucci (1988). In this coordinate the stellar matter
is closely self-similar, and the different shapes of the RC's curves
are mainly due to the $M_{vir}-M_D$ relation. In the space defined
by normalized circular velocity - dark radius - halo mass, spirals
do not occupy random positions, but a well defined plane of very
small thickness. We clearly see that, by filling only less than
$10^{-3}\%$ of the available volume, the available kinematics of
spiral galaxies defines the Universal Rotation Curve. Let us notice
that, in principle, theories of the formation of spirals do not
trivially imply the existence of such a surface that underlies the
occurrence of a strong dark - luminous coupling.

\begin{figure}
\centerline{\psfig{figure=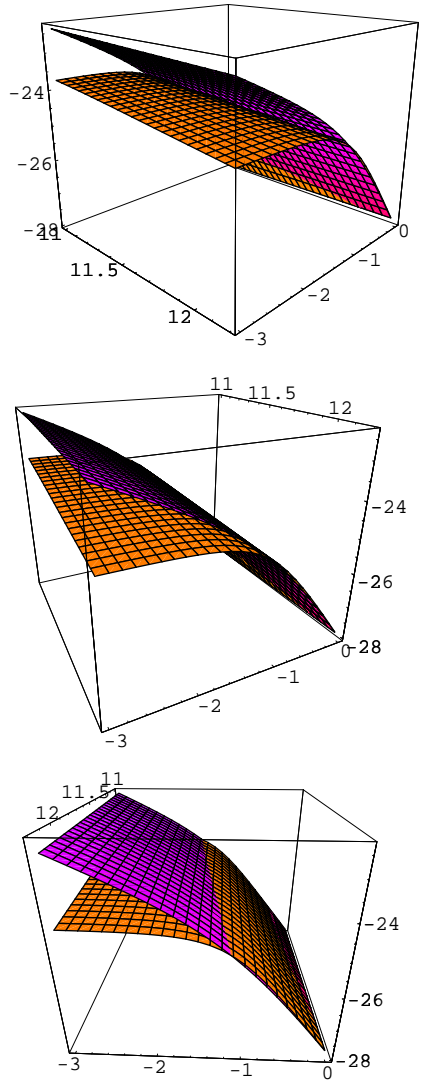,height=14.5cm}} \caption{The
URC halo density  vs. the  NFW halo density  of objects of the same
mass,  as a function of normalized radius and virial mass. The axes
labels are  $x $, $log M_{vir }/M_\odot$ and $log (\rho/ (g~
cm^{-3}))  $ } \vskip-.5truecm
\end{figure}

We now show the URC DM density distribution. In Fig.~6 we show it as
a function of $x$ and $M_{vir}$. For $x < 10^{-1}$ the well known
core-cusp discrepancy emerges, i.e. the DM density of actual halos
around spirals is about one order of magnitude smaller and radially
much  more constant than the NFW predictions. At $x > 0.4$, for a
concentration parameter $c = 13 (M_{vir}/10^{12}\
M_{\odot})^{-0.13}$, the observed halo densities are consistent with
the NFW predictions for halos of the same virial mass.
 Note that this is a direct test: for halos with density profiles at $x > 0.5$
very different from the Burkert or the NFW profiles, Eq.~9b does not
have solution.

More specifically, let us constrain the analytical
form of the outer DM distribution. For $2\leq r/r_0 \leq 18 $, the
following approximation for the Burkert and NFW profile holds
($y\equiv r/r_0$, $\epsilon = 0 $)
$$
V_{URCH}(y)=  V_{URCH}(3.24)\frac{2.06\,y^{0.86}}{1.59 + y^{1.19
+\epsilon}} ~.
\eqno(11)
$$
Let us suppose that the actual {\it outer} DM velocity {\it profile}
is different from the Burkert/NFW given by Eq.~(11), i.e. $\epsilon
\neq 0$.  Then in Fig 7 we show that, even assuming large uncertainties in
$V_{vir}$, in order to match both $V(R_l)$ and $V_{vir}$, we must
have $\epsilon < 0.1$. This is a first direct  support for the Burkert and
the NFW density law to be able to represent the outer regions ($0.3
\leq r/R_{vir} \leq 1$) of DM galaxy halos.

Notice that weak-lensing shear fields, at several hundreds kpc from
the galaxy centers, are found {\it compatible} with the predictions
of the NFW density profile, but cannot exclude non-NFW profiles (
Kleinheinrich  ., et al., 2006 and references therein).

\section{Discussion and conclusions}

In this paper, we have built the Universal Rotation Curve of spiral
galaxies by means of kinematical and photometric data. We physically
extended the URC, established for the inner region of galaxies in
PSS out to $R_{vir}$ and have been able to employ the virial mass
$M_{vir}$ as the parameter that characterizes spiral galaxies, and
the virial radius $R_{vir}$ as a unit of measure for the radial
coordinate. This URC is meant to be the {\it observational}
counterpart of the NFW rotation curve, emerging from cosmological
simulations performed in the CDM scenario. The URC yields the
gravitational potential at any radius and it allows to link the
local properties in the inner luminous regions with the global
properties of the DM halos.

DM halos have one (and likely just one) characteristic length scale,
$r_0 \propto M_{vir}^{0.6}$, which it is not {\it naturally} present
in current scenarios of galaxy formation. Thus, they do not show any
sign of an inner cuspy region of size $r_s\propto M_{vir}^{0.4}$.
The halo velocity contribution  $V_{URCH} $ rises with radius like a
solid body at $r\sim 0$, decelerates to reach a maximum at
$3.24~r_0$ from where it start to slowly decrease out to $R_{vir}$
with a slope that it is consistent with that of the NFW halos. The
main significance of the URC concerns the full mass distribution
(MD). First, it is possible to immediately exclude the following
scenarios (and combinations of them): i) individual behavior, every
object has its own MD; ii) unique behavior, every object has almost
the same MD. Instead, the MD in spirals shows a remarkable {\it
mass-dependent} systematics: both the dark and the stellar matter
are distributed according to {\it profiles} that are functions of
the total mass $M_{vir}$ (see Fig.~8). Finally, the DM halo becomes
the dominant mass component in galaxies at different radii,
according to the galaxy mass: from $\sim 10^{-2} R_{vir}$ for the
lowest masses, to $\sim 10^{-1} R_{vir}$ for the highest ones.

We write the equilibrium velocity of the halos around
spirals as the following approximation of the relations in the
previous Sect.:
$$
V_{URCH}= A(M_{vir}) x^{-1/2} \Big\{ ln \Big[ 1 + \gamma(M_{vir}) x
\Big] - \tan^{-1} \Big[ \gamma(M_{vir}) x \Big] +{1\over {2}} ln
\Big[ 1 + \gamma(M_{vir}) x^2 \Big] \Big\}^{0.5}
$$
with $A(M_{vir}) = 0.406 + 1.08 \ log [M_{vir}/(10^{11} \rm
M_\odot)] - 0.688 \ \{log [M_{vir}/(10^{11} \rm M_\odot)]\}^2 ) +
0.766 \ \{log [M_{vir}/(10^{11} \rm M_\odot)]\}^3$ and
$\gamma(M_{vir}) = 26.78 \ [M_{vir}/ (10^{11} \rm
M_\odot)]^{-0.246}$. This is the observational counterparts of
$N$-body outcomes. 

A Mathematica code for the figures in this paper is available at:
http://www.novicosmo.org/salucci.asp.

\begin{figure}
\vskip -.5truecm
\centerline{\psfig{figure=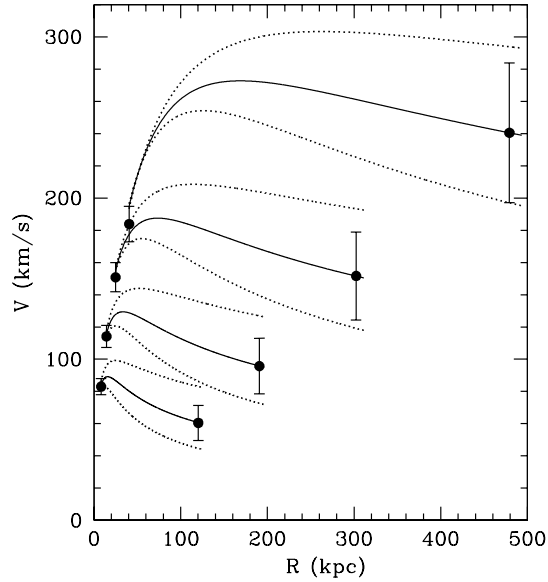,width=9.0cm}}
\vskip -.5truecm
\caption{Halo velocities at $R_l$ and $R_{vir}$
({\it filled circles}) vs. the URC-halo velocity, given by Eqs.~(11)
({\it solid line}) and vs. velocity profiles with average
logarithmic slope steeper or shallower by an amount $\epsilon= 0.1$
({\it dashed line}).}
\end{figure}

\begin{figure}
\vskip -3.5truecm
\centerline{\psfig{figure=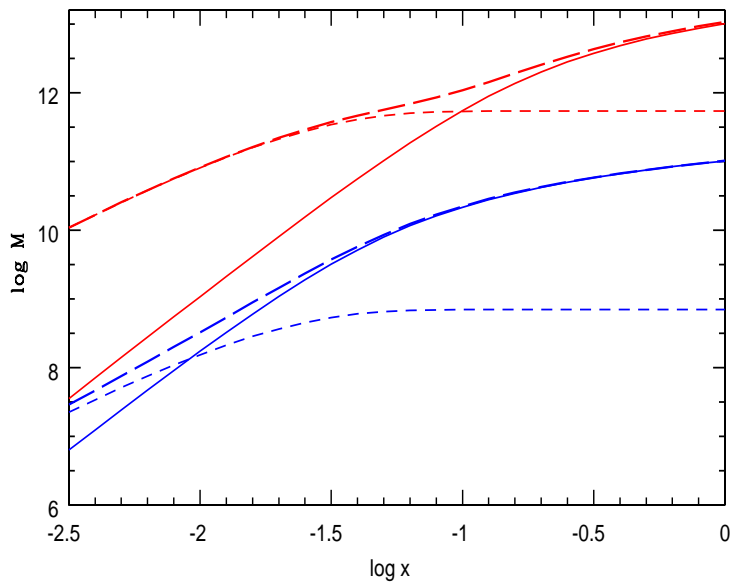,width=11.0cm,height=13.5cm}}
\caption{The dark halo (solid line), the disk (short dashed line),
and the total (long dashed) URC mass profile for reference masses of
$10^{11} \rm M_\odot$ and $10^{13} \rm M_\odot$. The distribution
for the intermediate  halo masses can be derived from section 2 }

\end{figure}

\section{Appendix}

In this Appendix we discuss  the observational  evidence  for the
URC claim, the  nature and the implications of which 
it is worth to  clarify. The paradigm states that, when binned by luminosity,
the RCs form a set of smooth, low-scatter synthetic curves, whose
profiles and amplitudes are strong functions of the luminosity bin.
\footnote{The  {\it analytical form}  of the   URC is  built by
assuming  reasonable  disk-halo  velocity profiles, with three
free parameters ($V(R_{opt}), a, \beta$) that are obtained by
$\chi^2$-fitting the sinthetic curves.}. Furthermore, the URC
paradigm implies: i) rotation velocity  slopes \textit{vs.} rotation
velocity amplitudes relationships (see fig (2) and (3)  of PSS) and
ii)  a set of relations (Radial Tully Fisher relationship) holding at
different radii $x$, defined as:
$$
log\  V(x)=a_x M + b_x
$$
where $x\equiv R/R_D$ and $a_x$ and $b_x$  are the fitting  
parameters, and $M$ is the galaxy  magnitude (Persic and Salucci 1991).

Evidence for the URC claim and/or its above implications comes from: 
a) detailed analyzes  of  independent samples:  
Catinella et al 2006, (2200 RC's, see their Fig. 12), Swaters 1999,
(60 extended RC's, see Chapter 4);  b) independent analyzes  of the PSS sample:  Rhee (1996), Roscoe (1999);
c) the  finding of a  very tight  RTF in PSS and other three different samples 
Willick (1997, see below),  Yegorova et al. (2007).

The  claim  has been also  tested   by  comparing the  RCs of two
samples of spirals (Courteau, 1998,  131 objects;   Verheijen, 1997,
30 objects)  with the circular velocities predicted by the  URC$_0$,
once  that  the values of  galaxy  luminosity and disk length-scale
are inserted in it.  The face-value  result  of the test:   2/3 of
the RCs are  in pretty good agreement  with the universal curve,
while  1/3 show some disagreement, indicates that  the  URC   is a
useful tool    to  investigate the systematics of  the mass
distribution in spirals,  but also it  questions  about  its
universality. However, while some of this    disagreement may
reflect an  inefficiency  of the URC$_0$ to reproduce the RCs, the
actual performance of the URC is better than it is claimed. In fact,
spurious  data vs  predictions  disagreements  are created in
performing this  test and precisely when they insert  in   URC$_0$   the   values of $L_B$
and $R_D$,  affected by   (occasionally large)  observational errors. 
 By taking into account this effect the
URC$_0$ success rate reaches $80\%$ and more.

Willick (1998) found, by studying a large  sample of RCs, a radial
variation  of the scatter of the inverse RTF defined above and he
interpreted   it as  an evidence against the URC. Let us show that
this argument  is  incorrect and that {\it au contraire} the
properties of the RTF support the URC paradigm. The
increase/decrease of the  scatter found is very small (Willick,
1998): the scatter ranges from  0.065 dex (at $2 R_D$ )  to  0.080
dex  (at  $0.5 R_D$ and at  $3 R_D$) and it implies,  if  totally
intrinsic, a prediction error  in $log\ V(x)$ of
$(0.08^2-0.065^2)^{0.5} = 0.04\  dex$. Moreover,    some of the
scatter increase/decrease  is due to the  larger  random
observational errors present in the outermost  measurements; in fact, a
refined analysis of the issue (Yegorova et al 2007) finds a
smaller predicting error for three large sample of spirals.
Therefore, from the RTF we have that, in the region considered, the
luminosity  {\it statistically}  predicts the circular velocity at
any radius and in any galaxy within an error of 5\% - 10\% , a
quantity much smaller than the variations of the latter  {\it in
each} galaxy and {\it among} galaxies.

\section*{acknowledgements}
We thank L. Danese for helpful discussions, and the referee for useful comments.

 \section{References}

\noindent

\bibitem{} 
Baldry, I.K.,  et al.  2004, ApJ, 600, 681  
 
\bibitem{}
Bell, E.F., McIntosh, D.H., Katz, N., \& Weinberg, M.D.
2003, ApJS, 149, 289  

\bibitem{} 
Borriello, A., Salucci, P. 2001, MNRAS, 323, 285  

\bibitem{} 
Bosma, A. 1981, AJ, 86, 1791  

\bibitem{}
 Broeils, A.H. 1992a, A\&A, 256, 19  

\bibitem{}
 Broeils, A.H. 1992b, Ph.D. thesis, Groningen University  

\bibitem{} 
Catinella, B., Giovanelli R.,  Haynes, M.P., 2006,  ApJ. 640, 751  

\bibitem{}
 Courteau S., 1997, AJ, 114, 2402  

\bibitem{} 
Donato, F., Gentile, G., Salucci, P., 2004, MNRAS, 353, 17  

\bibitem{}
 Eke, V.R., Cole, S., \& Frenk, C.S., 1996, MNRAS 282,263  

\bibitem{} 
Freeman, K.C. 1970, ApJ, 160, 811  

\bibitem{} 
Gentile, G., Burkert, A., Salucci, P., Klein, U., Walter, F.
2005, ApJ, 634, L145  

\bibitem{} 
 Gentile G., Salucci P., Klein U., Vergani D., Kalberla P., 2004, MNRAS, 351, 903  

\bibitem{} 
 Gentile G., Salucci P., Klein U., Granato  G.L., 2007, MNRAS, 375, 199  

\bibitem{} 
Kleinheinrich M., et al., 2006, A\&A, 455,441  
 
\bibitem{} 
Mo, H.J., Mao, S., White, S.D.M. 1998, MNRAS, 295, 319  

\bibitem{} 
Noordermeer, E, 2007, MNRAS, in press,  astro-ph/0701731  

\bibitem{} 
Navarro, J.F., Frenk, C.S., \& White, S.D.M. 1996, ApJ, 462, 563  

\bibitem{} 
Persic, M., Salucci, P. 1988, MNRAS, 234, 131 (PS88)  

\bibitem{} 
 Persic M., Salucci P., 1990, MNRAS, 245, 577  

\bibitem{} 
Persic, M., Salucci, P. 1991, ApJ, 368, 60 (PS91)  

\bibitem{} 
Persic, M., Salucci, P. 1995, ApJS, 99, 501 (PS95)  

\bibitem{} 
Persic, M., Salucci, P., Stel, F. 1996,  MNRAS, 281, 27

\bibitem{} 
Rhee, M-H 1997, Thesis, Groningen University. 

\bibitem{}
 Roscoe, D. F. 1999, A\&A {\bf 343}, 788  

\bibitem{} 
Rubin, V.C., Ford, W.K., Jr., Thonnard, N. 1980, ApJ, 238, 471  

\bibitem{}
 Rubin, V.C., Ford, W.K., Jr., Thonnard, N., Burstein, D., 1982,
ApJ, 261, 439  

\bibitem{} 
Rubin V. C., Burstein D., Ford W.K., Jr., Thonnard N., 1985, ApJ, 289, 81  

\bibitem {} 
Salucci P., Burkert A., 2000, ApJ, 537, L9  

\bibitem{} 
 Salucci P., Persic M., 1997, ASPC, 117, 1  
 
\bibitem{} 
Shankar, F., Lapi, A., Salucci, P., De Zotti, G., Danese, L.
2006, ApJ, 643, 14  

\bibitem{}
 Simon, J. D., Bolatto, A. D., Leroy, A., Blitz, L., Gates, E. L.
2005, ApJ, 621, 757  

\bibitem{}
 Swaters, R.A., Madore, B.F., van den Bosch, F.C., Balcells, M.
2003, ApJ, 583, 732  

\bibitem{} 
Swaters, 1999, Ph.D. Thesis, Groningen University  

\bibitem{} 
Tonini, C., Lapi, A. \& Salucci, P. 2006, ApJ, in press  

\bibitem{}
 Tonini, C., Lapi, A., Shankar, F. \& Salucci, P. 2006,
ApJ, 638, L13  

\bibitem{} 
van den Bosch, F.C., Swaters, R.A., 2001 MNRAS, 325, 1017

\bibitem{}  
Verheijen, M,  1997,   Ph.D. Thesis, Groningen University  

\bibitem{}
 Yegorova, I. et al, 2007, MNRAS in press, astro-ph/0612434    
 
\bibitem{} 
Weldrake, D.T.F., de Blok, W. J. G., Walter, F. 2003,
MNRAS, 340, 12  

\bibitem{}   Willick J.A., 1999, ApJ, 516, 47

\end{document}